\documentstyle[prb,aps,multicol,epsfig]{revtex}
\begin{document}
\title{Generalized hole-particle transformations and spin reflection positivity in multi-orbital systems}
\author{Mario Cuoco and Canio Noce}
\address{I.N.F.M. -Unit\`a di Salerno and Dipartimento di
Fisica ''E.R. Caianiello'', \\ Universit\`a di Salerno I-84081
Baronissi (Salerno), ITALY}
\maketitle
\begin{abstract}
{We propose a scheme combining spin reflection positivity and
generalized hole-particle and orbital transformations to
characterize the symmetry properties of the ground state for some
correlated electron models on bipartite lattices. In particular,
we rigorously determine at half-filling and for different regions
of the parameter space the spin, orbital and $\eta$ pairing
pseudospin of the ground state of generalized two-orbital Hubbard
models which include the Hund's rule coupling.}
\end{abstract}
%\PACS{
%      {75.10.Lp}{Magnetic properties and materials}   \and
%      {71.27.+a}{Electronic structure}
%     } % end of PACS codes
%} %end of abstract
%
%CODE: 75.30.Mb
%Magnetic properties and materials : Intrinsic properties of
%magnetically ordered materials :[Valence fluctuation, Kondo
%lattice, and heavy-fermion phenomena]
%
%CODE: 71.27.+a Electronic
%structure : Strongly correlated electron systems; heavy fermions
%
%CODE: 75.10.Dg Magnetic properties and materials : General theory
%and models of magnetic ordering :[Crystal-field theory and spin
%Hamiltonians]
%
\begin{multicols}{2}
\section{Introduction}
\label{intro} Since the discovery of high temperature
superconductors, a renewed interest has increased in models
describing strongly correlated electrons. As the simplest one the
Hubbard model has been intensively studied within the analysis of
strongly correlated electron systems. Apart from it, various
generalized Hubbard models have also attracted considerable
interest.{\cite{hubext}} Indeed, by including some additional
interaction terms into the Hubbard Hamiltonian, either the
magnetic ordering phases or the superconducting phases may be
stabilized in appropriate regions of the parameter
space.{\cite{deboer}} The introduction in these models of orbital
degrees of freedom{\cite{roth}} enriches the phase diagram of the
model, because the interplay of orbital and spin dynamics can
yield a series of novel physical phenomena, such as unconventional
metal-insulator transition, novel quantum spin and orbital ordered
states, and transport property anomalies.\cite{tokura}

Generally, all of these models retain only part of the
electron-electron interactions which arise as a result of the
repulsion between two electrons having opposite spins and located
at the same lattice site. In spite of this crude assumption, these
models correspond to a many-body problem and in general cannot be
exactly solved. Exact solutions are of great importance, since in
some cases the errors introduced by the approximations may
dominate the results to such an extent that one might end up with
an incorrect description of the phenomenon under study.
Considering the exact results with respect to the dimensionality
of the model, we know that most of them have been derived in two
limiting case: either in one dimension or in the infinity
dimensions. For instance, the exact solution of the Hubbard model
was given in the one dimensional case by means of the Bethe ansatz
by Lieb and Wu{\cite{libwu}}. The other class of exact solutions
belongs to the other limiting case, i.e. d=$\infty $, where the
dynamical mean-field approximation becomes exact.{\cite{vol}} The
situation gets more complicated as physically interesting lower
dimensional cases are considered. d$>$1 rules out the
applicability of the well-established Bethe ansatz approach, while
mean-field-like descriptions lead to qualitatively or
quantitatively incorrect conclusions, because the effects of
spatial and dynamical fluctuations are not properly taken into
account. Nevertheless, exact results holding in any dimensions are
available. We refer to, for instance, the Lieb
theorem{\cite{lib}}, the flat-band ferromagnetism{\cite{milk}},
and the Nagaoka theorem.{\cite{naga}} However, for extended
versions of the Hubbard model exact results are still
rare.{\cite{su}},{\cite{kor}}

In this paper we rigorously prove some exact results for a
generalized Hubbard model. Namely, we consider a two-orbital
Hubbard Hamiltonian which incorporates also the Hund's rule
coupling. We notice that the study of relevant physical systems
requires the use of such kind of model whose minimal constituents
are the hopping term between different, or same, orbitals, and the
on-site Coulomb and exchange interactions. This is for instance,
to report few of them, the case of non-metallic
Cu$^{2+}$compounds{\cite{feld}} (d$^9$ configuration, one hole in
the two degenerate $e_g$-orbitals), low spin
Ni$^{3+}${\cite{AMOL}}(d$^7$ configuration, one electron in
$e_g$-orbitals), as well as Mn$^{3+}$ and Cr$^{2+}${\cite{AMOL}}
ions (high-spin d$^4$ configuration, one $e_g$ electron), and in
the Ru$^{4+}$ ions{\cite {maeno}} (d$^4$ configuration, two holes
in $t_{2g}$ -orbitals).

Here, our aim is to determine rigorously the symmetry properties
of the above mentioned Hubbard Hamiltonian by investigating which
values of the spin, orbital and $\eta$ pairing pseudospin quantum
numbers occur in the ground state. A physical interpretation and
discussion of the outcome is also presented. The following
conditions are assumed for the validity of the results obtained
below: i) the number of electrons is equal to twice number of
sites in the lattice; ii) the lattice satisfies the connectivity
condition; iii) the hopping amplitude is different from zero only
for charge transfer between orbitals of the same type.
\\ The scheme of the proof develops into three preliminary steps
which are based on the use of the property of reflection
positivity of the Hamiltonian in an opportune range of parameters
and on the application of continuity arguments derived by the
uniqueness of the ground state. Furthermore, the link introduced
by some unitary transformations between the ground states in
different regions of the space of parameters, allows to recover
all the missing information about their symmetry properties.

\section{The model}
\label{sec:1} The explicit form of the Hamiltonian for a double
orbitally degenerate Hubbard model on a connected bipartite
lattice $\Gamma$ is written as:
\begin{equation}
H=H_t+H_{int}, \label{eq:ham}
\end{equation}
where
\begin{equation}
H_t=-t\sum_{<ij>,\lambda ,\sigma }d_{i\lambda \sigma }^{\dagger
}d_{j\lambda \sigma }
\end{equation}
\[
H_{int}=U_0 \sum_{i,\lambda }n_{i\lambda \uparrow }n_{i\lambda
\downarrow}+U \sum_{i,\sigma}n_{i1 \sigma }n_{i2 \bar{\sigma}}+
\]
\begin{equation}
(U-J)\sum_{i,\sigma}n_{i1 \sigma }n_{i2 \sigma}- J\sum_{i,\sigma}
d_{i 1 \sigma}^{\dagger}d_{i 1 \bar{\sigma} }d_{i 2
\bar{\sigma}}^{\dagger }d_{i 2 \sigma}
\end{equation}

\noindent Here $d_{i\lambda \sigma }^{\dagger }$ is the creation
operator of correlated electrons with spin $\sigma $ at site $i$
on orbital $\lambda $ (=1,2), respectively and
$n_{i\lambda\sigma}$ is the number operator for the electrons on
the i site and $\lambda$ orbital. Moreover, we have used the
simplified notation $\bar{\sigma}=-\sigma$. The parameter $t$ is
the hopping matrix between nearest-neighbor sites, and we assume
that the $\lambda$ orbitals are not mixed by the hopping; $U_0$,
$U$ and $J$ stand for the intra-orbital, inter-orbital Coulomb and
Hund's rule exchange interaction, respectively.

The results presented below are valid for any choice of the
parameters in the Hamiltonian. Nevertheless, since the two
orbitals are equivalent and can be interchanged by a properly
chosen canonical transformation, we impose an additional condition
on the set of the parameters, that is $U_0=U+J$.\cite{AMOL,grif}
Hereafter, we will use this relation together with the condition
that $U$ and $J$ are positive. When will be needed, the sign of
the Coulomb and Hund coupling will appear explicitly. Hence
$H_{int}$ becomes,
\[
H_{int}=(U+J)\sum_{i,\lambda }n_{i\lambda \uparrow }n_{i\lambda
\downarrow}+U \sum_{i,\sigma}n_{i1 \sigma }n_{i2 \bar{\sigma}}+
\]
\begin{equation}
(U-J)\sum_{i,\sigma}n_{i1 \sigma }n_{i2 \sigma} - J\sum_{i,\sigma}
d_{i 1 \sigma}^{\dagger}d_{i 1 \bar{\sigma} }d_{i 2
\bar{\sigma}}^{\dagger }d_{i 2 \sigma}
\end{equation}
Let us introduce the following operators:
\begin{eqnarray}
{\bf S}&=&\frac 12\sum_{i,\sigma,\sigma ^{\prime
}\lambda}d_{i\lambda \sigma }^{\dagger }({\bf \sigma })_{\sigma
\sigma ^{\prime }}d_{i\lambda \sigma ^{\prime }}\\
{\bf T}&=&\frac 12\sum_{i,\sigma,\lambda,\lambda ^{\prime } }d_{i\lambda \sigma }^{\dagger }({\bf \sigma }%
)_{\lambda \lambda ^{\prime }}d_{i\lambda ^{\prime }\sigma }\\
{\mathbf \eta}&=&\frac 12\sum_{i,\sigma,\sigma ^{\prime }\lambda}
\varepsilon(i) d_{i\lambda \sigma }^{\dagger }({\bf \sigma
})_{\sigma \sigma ^{\prime }}d_{i\lambda \sigma ^{\prime
}}^{\dagger }
\end{eqnarray}

\noindent where ${\bf \sigma }$ are the Pauli matrices and
$\varepsilon(i)=\pm1$ depending to which of the two subparts of
the bipartite lattice the site $i$ belongs.

\noindent Here ${\bf S}$ is the usual total spin operator, ${\bf
T}$ is the pseudospin operator and ${\bf \eta}$ is the so called
pairing operator introduced by Yang \cite{Yang92}, extended to the
case of two types of electrons. The ${\bf T}$ operator has
properties that are exactly analogous to the properties of the
usual one-half spin operator.
Indeed, its third component at each site assumes the values $\frac 12$ and $%
-\frac 12$ corresponding to the occupied orbitals $\lambda =1$ and
$2$, respectively. Besides, $T_i^{+}$ takes a fermion in the
orbital 2, at the lattice site i, and transports it to the orbital
1 located at the same lattice point. Obviously, $T_i^{-}$
corresponds to the opposite process.
\\
As for ${\bf S}$ and ${\bf T}$, the ${\bf \eta}$ operators
generate another $SU(2)$ algebra which has as base configurations
those with one orbital doubly occupied or empty on each site.

In this model, the total spin operator ${\bf S}$ commutes with the
Hamiltonian $\left[ {\bf S,}H\right] =0$ and is a good quantum
number. Another symmetry is that one related to the orbital degree
of freedom as defined by ${\bf T}$. The square of the total
orbital pseudospin operator ${\bf T}$ and its third component
$T^z$ commute with the Hamiltonian $H$, since there is no hopping
between different orbitals in $H_t$. On the other hand, the square
of the total $\eta$ pseudospin operator and its third component
$\eta^z$ also commute with the Hamiltonian. Hence, these relations
imply that these are conserved quantities and that the eigenstates
of $H$ can be labelled as follows:
\begin{center}
$|\bullet >=|E,S,S_z,T,T_z,\eta,\eta_z>.$
\end{center}
\noindent We also notice that $T^{-}$ and $T^{+}$ commutes with
$H$, namely the $T-$ multiplets are degenerate in energy. This
property is not valid for the $\eta^{+}$ and $\eta^{-}$ operators.

It is important pointing out that these algebras are not
independent each-other and can be related by means of extended
hole-particle and orbital type transformations.

Let us consider the following unitary transformation ${\bf G}$:
\begin{equation}
{\bf G}d_{i1\uparrow }{\bf G}^{-1}=d_{i1\uparrow }
\end{equation}
\begin{equation}
{\bf G}d_{i1\downarrow }{\bf G}^{-1}=d_{i2\uparrow }
\end{equation}
\begin{equation}
{\bf G}d_{i2\uparrow }{\bf G}^{-1}=d_{i1\downarrow }
\end{equation}
\begin{equation}
{\bf G}d_{i2\downarrow }{\bf G}^{-1}=d_{i2\downarrow }.
\end{equation}
\noindent Under this transformation one has
\begin{equation}
{\bf G}\bf {S} {\bf G^{-1}=}\bf {T},
\end{equation}
\noindent and the Hamiltonian H is transformed as follows:
\begin{equation}
{\bf G}H{\bf G^{-1}}{{\bf =}}{\bf G}(H_t+H_{int}){\bf G^{-1}}%
{{\bf =}}\tilde{H_G}.
\end{equation}
\noindent Here $\widetilde{H}_G=H(t,J\Rightarrow -J),$ i.e. $
\widetilde{H}_G$ can be obtained from $H$ replacing $J$ with $-J$,
implying that $\widetilde{H}_G$ and $H$ are related to each other.

There is also another transformation that will appear very useful
later for the determination of the symmetry properties of the
ground state. This is the hole-particle transformation of the base
operators of creation and annihilation:
\begin{equation}
{\bf R}d_{i1\uparrow }{\bf R}^{-1}=d_{i1\uparrow }
\end{equation}
\begin{equation}
{\bf R}d_{i1\downarrow }{\bf R}^{-1}=\varepsilon(i)
d_{i1\downarrow}^{\dagger}
\end{equation}
\begin{equation}
{\bf R}d_{i2\uparrow }{\bf R}^{-1}=d_{i2\uparrow }
\end{equation}
\begin{equation}
{\bf R}d_{i2\downarrow }{\bf R}^{-1}=\varepsilon(i)
d_{i2\downarrow }^{\dagger}.
\end{equation}

\noindent Under this transformation one has
\begin{equation}
{\bf R}\bf {S} {\bf R^{-1}=}\bf {\eta},
\end{equation}
\noindent and the Hamiltonian $H$ is transformed as follows:
\begin{equation}
{\bf R}H{\bf R^{-1}}{{\bf =}}{\bf R}(H_t+H_{int}){\bf R^{-1}}%
{{\bf =}}\widetilde{H}_R.
\end{equation}
\noindent Here $\widetilde{H}_R=H(t,J\Rightarrow -J,U\Rightarrow
-U),$ i.e. $ \widetilde{H}_R$ can be obtained from $H$ replacing
$J$ with $-J$ and $U$ with $-U$, implying that $\widetilde{H}_R$
and $H$ are related to each other.

\begin{figure}
\centerline{\epsfxsize=7.5cm \epsfbox{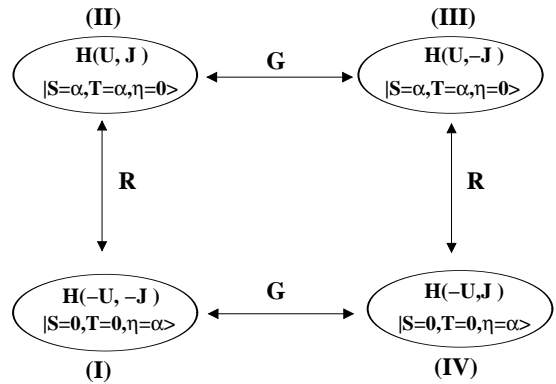}} \caption{A
sketch of the links introduced by ${\bf R}$ and ${\bf G}$ between
the Hamiltonian and their respective ground states as discussed in
step $I-III$. Each box contains now the complete information about
the quantum numbers of the ground state in the different regions
of parameter space.} \label{fig:schema1}
\end{figure}

Our purpose is to determine rigorously the symmetry properties of
the two-band Hubbard model as defined above, by investigating
which values of the spin, orbital and $\eta$
pairing pseudospin quantum numbers occur in the ground state. \\
As we have pointed out in the Introduction, the proof is based on
three preliminary steps, followed by the application of the ${\bf
R}$ and ${\bf G}$ transformations to connect the ground states in
different regions of the space of parameters. We will describe how
this scheme may allow to get rigorous information on the symmetry
property of the ground state
%CAMBIA
even where the of spin reflection positivity is not directly
applicable, thus yielding a more general description of the ground
state symmetry properties.
%CAMBIA
\\
{\bf{Step I}}: \\ \noindent{Let consider firstly the case derived
by applying the transformation $J\rightarrow -J$ and $U\rightarrow
-U$ on $H(U,J)$ which changes $H(U,J)\rightarrow H(-U,-J)$. We
remind that this transformation is realized by the application of
the unitary transformation ${\bf R}$ on $H(U,J)$. Having this
modified Hamiltonian the property of spin reflection positivity,
it is possible to state\cite{lib} that the ground state $|G
\rangle_{I}$ of $H(-U,-J)$, for any positive value of $U$ and $J$,
is unique and has total spin $S=0$.
\\ Moreover, for this case it is possible to determine the value of the orbital
pseudospin quantum number, by looking at the relevant
configurations which contribute to the ground state at values of
$|U|/t\gg1$. As $-U$ is large and negative, one ends up with
states which contain only configurations with two electrons
occupying the same orbital on one site. The total orbital
pseudospin for this kind of situation is $T=0$, since it is
identically zero on any configuration with two electrons in the
same orbital state, i.e. it is a singlet in the orbital space. It
is worth pointing out that the previous statement can be
generalized only at the condition that the ground state is
non-degenerate. This property permits the use of continuity like
arguments to extend the validity of the result from the extreme
large negative limit to any finite value of $-U$. }
%\end{itemize}
\\
{\bf{Step II}:}
\\
\noindent {Let now analyze the parameter case relative to the
Hamiltonian $H(U,J)$ which is the most relevant for systems of
physical interest. It is worth pointing out that the ground state
of $H(U,J)$ is unique as it is directly connected to that one of
$H(-U,-J)$ by the application of
the unitary transformation ${\bf R}$ defined in Eqs.14-19. \\
Moreover, by performing a perturbation up to second order in
$U/t$, it is possible to show that the low energy processes of
$H(U,J)$ are described by an effective Heisenberg model for spin
one half on an array composed by two sublattices $A$ and $B$
defined depending on the sign of the magnetic exchange between the
spin belonging to the two sublattices. The effective Hamiltonian
reads as follows:
\begin{eqnarray*}
H_{II}=J_1 \sum_{i,j} ( S_{i(A)} S_{j(A)}+ S_{i(B)} S_{j(B)}) +J_2
\sum_{i,j} S_{i(A)} S_{j(B)}
\end{eqnarray*}
\noindent where $J_1=-J$, i.e. the direct Hund coupling and $J_2=4
t^2/(U+J)$. For this Hamiltonian, it has been proved \cite{lieb62}
that the ground state is unique, apart from the usual $SU(2)$
degeneracy, and it has total spin $S=\frac{1}{2}(|A-B|)$, where
$A(B)$ are the total number of lattice sites belonging to the
sublattice $A(B)$, respectively. For building the mapping, one has
to consider an array where each orbital $\lambda_i$ on the generic
$j$ site is associated with an effective one, so that
$\frac{1}{2}(|A-B|)$ is equal to
$\frac{1}{2}|(N_{\lambda_1}(A)-N_{\lambda_1}(B))+
(N_{\lambda_2}(A)-N_{\lambda_2}(B))|$, $N_{\lambda_i}(K)$ being
the number of $\lambda_i$ orbitals belonging to the sublattice
$K$. \\
It is important to stress that it is again the uniqueness property
of the ground state of $H_{II}$ and $H(U,J)$ that can permit the
use of continuity arguments to state that the ground state $|G
\rangle_{II}$ of $H(U,J)$ is nondegenerate except for the $SU(2)$
symmetry, and it has total spin $S=\frac{1}{2}(|A-B|)$. }
\\
{\bf{Step III}:}
\\
\noindent {Finally, let analyze the situation obtained by applying
to the Hamiltonian $H(U,J)$ the transformation ${\bf R}$ which
changes $J\rightarrow(-J)$. By means of this transformation,
$H(U,J)$ is modified in $H(U,-J)$ where the Hund coupling assumes
the form of antiferromagnetic instead than ferromagnetic type
interaction.
\\ As it has been performed in the step II, by means of an expansion up
to the second order in $U/t$, the search of the symmetry
properties of the ground state of $H(U,-J)$ may be carried into
the analysis of the ground state for an effective Heisenberg model
with two kinds of exchanges.
\\
The Hamiltonian for the lower energy processes is given by the
following expression,
\begin{eqnarray*}
H_{III}=J_1 \sum_{i,j} ( S_{i(A)} S_{j(A)}+ S_{i(B)} S_{j(B)})
+J_2 \sum_{i,j} S_{i(A)} S_{j(B)}
\end{eqnarray*}
\noindent where $J_1=-J$, i.e. the direct Hund coupling with
inverted sign and $J_2=4 t^2/(U+J)$. As for the previous case, by
using the arguments of Lieb\cite{lieb62}, one can show that the
ground state is unique except for the $SU(2)$ degeneracy and has
total spin $S=\frac{1}{2}(|A-B|)$. Following the same procedure of
step I, one can state by means of continuity arguments that the
ground state $|G \rangle_{III}$ of $H(U,-J)$ is unique and has
total spin $S=\frac{1}{2}(|A-B|)$. Let remind that the value of
$A(B)$ is related to the number of orbitals belonging to the
sublattice $A(B)$, respectively.}\\
%\end{itemize}
%
\begin{figure}
\centerline{\epsfxsize=7.5cm \epsfbox{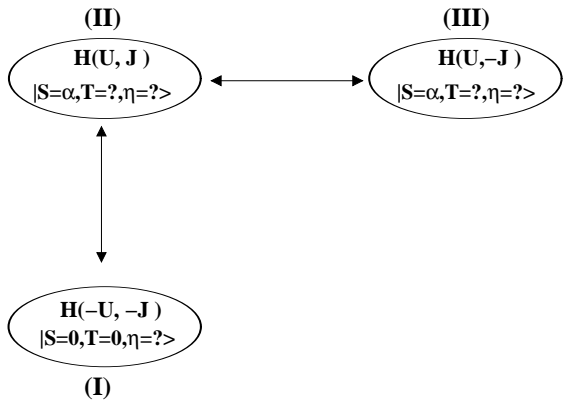}} \caption{Scheme
of the symmetry properties of the ground state of the two-band
Hubbard model for different regions of the parameter space as
derived from the procedure of step $I-III$.} \label{fig:schema}
\end{figure}
As a summary in Fig. \ref{fig:schema} it has been reported the
main findings out of the procedure performed above. Each box
contains the information about the quantum numbers of the ground
state and to which region of the parameter space it refers. The
ket $|S=\alpha,T=?,\eta=?\rangle$ stands for the ground state,
whose $S$,$T$, and $\eta$ are the values of the total spin,
orbital and $\eta$ pairing pseudospin quantum numbers with
$\alpha$ given by $\frac{1}{2}(|A-B|)$.\\ As one can see in Fig.
\ref{fig:schema}, there are some quantum numbers which cannot be
derived from the analysis presented in step $I-III$, and for this
reason the missing values have been indicated with a question
mark.
At this point, it is crucial the use of the transformations ${\bf
G}$ and ${\bf R}$ to link the different ground states and the
correspondent quantum numbers. Indeed, taking advantage of these
symmetry transformations, one can complete and extend the scenario
as in Fig. \ref{fig:schema} by
building up the scheme as it is presented in Fig. \ref{fig:schema1}.\\

To figure out how to determine the missing information on the
ground state, let remind that if we have a simultaneous eigenstate
of the operator ${\bf S}$ and of $H(U,J)$ with assigned spin
eigenvalue given by $s$, then the unitary operator ${\bf G}$
transforms this state into an eigenstate of $H(U,-J)$ and of ${\bf
T}$ whose pseudospin orbital eigenvalue is equal to $s$ too. Then,
one can proceed from case II to III and extract the information
that the ground states for $H(U,J)$ and $H(U,-J)$ have the same
value of the orbital pseudospin, i.e. $\alpha$, so that $|G
\rangle_{II}=|S=0,T=\alpha,\eta=?>$ and $|G
\rangle_{III}=|S=0,T=\alpha,\eta=?>$.
\\
In the same way, the unitary operator ${\bf R}$ is mapping an
eigenstate of the total spin and of the Hamiltonian $H(U,J)$ into
an eigenstate with equal total $\eta$ pairing pseudospin and of
$H(-U,-J)$.\\
Hence by linking via ${\bf R}$ the ground states of I and II whose
one knows the value of the total spin, it is possible to deduce
the pseudospin values which are now equal to $0$ and $\alpha$,
respectively. After that, we have full knowledge of the symmetry
character of the ground state in I and II as given by
$|G\rangle_{I}=|S=0,T=0,\eta=\alpha \rangle$ and
$|G\rangle_{II}=|S=\alpha,T=\alpha,\eta=0 \rangle$, though it is
still missing the information on the $\eta$ pseudospin in the
region III.
It comes natural at this point to enlarge the range of
investigation by moving to the case IV which corresponds to the
region of parameter space where $U$ is negative and $J$ is
positive as indicated by $H(-U,J)$, where a priori we do not have
any indication on the symmetry character of its ground state.
However, the region IV can be reached either via I or III so that
we can extract the complete information on all the quantum numbers
and cover the missing points in II and III. The path from I to IV
turns out to be useful for determining the spin and orbital
pseudospin values while the way to IV via III can yield the
information on the $\eta$ pairing pseudospin so to complete the
features for the ground state in these two regions, that is
$|G\rangle_{III}=|S=0,T=\alpha,\eta=0>$ and
$|G\rangle_{IV}=|S=0,T=0,\eta=\alpha>$, respectively.
\\
%CAMBIA
We would like to point out that there is an alternative way to
determine the total spin in the ground state which does not go
through the perturbative arguments used in steps II-III and that
it only requires the condition of semi-positive definiteness of
the ground state(s). The theorem which refers to this method has
been demonstrated and extensively discussed in Ref.\cite{shen98}
and it is based on the condition that a positive definite state
and a positive semi-definite state built on the same basis vector
space have a nonzero overlap. The key element of this proof is to
show that the ground state of the assigned Hamiltonian is positive
semi-definite and that the ground state of a related spin
Hamiltonian built on the same lattice and on the same Hilbert
space is positive definite. Hence the condition of zero overlap
between those states ensures that they have the same value of the
total spin. The relevance of this way of proceeding is that so
doing one does not have to prove the non-degeneracy of the ground
state which is fundamental in the proof that makes use of
perturbative like arguments. \\
For the case of Hamiltonian (\ref{eq:ham}), the proof of the
semi-positive definiteness of the ground state is not direct
because one has to use a set of complete and orthonormal basis
built by means of the operators ${\bf R}$ and ${\bf G}$. Indeed,
due to the property of spin reflection positivity and to the
filling chosen, one can introduce a set of complete and
orthonormal basis of the form $\{{\bf K}
|\phi_{\alpha}^{\uparrow}\rangle\otimes|\phi_{\beta}^{\downarrow}\rangle
\} $, where ${\bf K}={\bf R}({\bf G})$ and
$|\phi_{\alpha}^{\sigma}\rangle=\prod_{i,\lambda \epsilon \alpha}
d^{\dagger}_{i\lambda\sigma}$ with $\alpha$ being one of the
possible configurations. Then, it is possible to prove that states
obtained by an expansion on the previous basis as
$|W\rangle=\sum_{\alpha,\beta} W_{\alpha,\beta} {\bf K}
|\phi_{\alpha}^{\uparrow}\rangle\otimes|\phi_{\alpha}^{\downarrow}\rangle
$ have the properties of semi-positive definiteness, that is all
the eigenvalues of the matrix $W_{\alpha,\beta}$ are not less than
zero. \\
At this point it is possible to recover the results of steps
II-III by choosing a spin Hamiltonian of the form of $H_{II}$-
$H_{III}$ whose the ground state is positive definite and use the
condition of zero overlap to get the value of the total spin.
%CAMBIA
\\
Let consider now few comments about the physical content of the
symmetry character of the ground states in the regions I-IV. One
important observation is that $|G\rangle_{II}$ is a novel quantum
state which shows coexistence of unsaturated ferromagnetic spin
and orbital order on a lattice assumed that $A\neq B$, in the
sense that the ground state exhibits a value of the total spin and
orbital momentum which is different from zero and scales as the
total number of lattice sites. Moreover, as discussed in
Ref.\cite{shen93}, the state $|G\rangle_{I}$ and $|G\rangle_{IV}$
which have $\eta=\alpha$ can support ODLRO (off diagonal long
range order) as it concerns the long distances pair correlations.
It is striking that even in presence of positive Coulomb
interaction (i.e region IV) one can have an $\eta$ pairing in the
ground state. This can be understood considering that the pairing
for a multi-orbital system can come in a higher orbital momentum
state induced by the magnetic exchange so that the effect of the
local Coulomb repulsion is reduced.
\\
Another interesting physical property comes from the $\eta$
pairing symmetry of the ground state of the region II and III. Due
to the value assumed by $\eta$, it can be shown\cite{pu94,cuoco99}
that this state can support coexistence of charge density wave
(LRO) and superconductivity (ODLRO) in a phase which is usually
indicated as supersolid. The statement is true in the sense that
if LRO appears then by symmetry the state manifests also ODLRO and
viceversa.
\\
Indeed, it comes that due to the rotational invariance of the
ground state in the $\eta$ space, the Fourier transform of the
transverse correlation function at equal time $\eta^{+-}({\bf q})$
are proportional to the diagonal one $\eta^{z}({\bf q})$, thus if
one of the two has the property to stay finite for some value of
the momentum ${\bf q}$ in the thermodynamic limit, the same
happens for the other one.

\section{Conclusions}
In conclusion we have shown that the symmetry features of the
ground state at half filling of a generalized Hubbard model which
includes the Hund's rule coupling, can be obtained in a large
range of the paramater space, only by combining the property of
spin reflection positivity and the use of special unitary
transformations. In particular it has been possible to extract all
the most relevant symmetry characteristics related to conservation
laws of the Hamiltonian that are represented by the orbital, spin
and $\eta$ pairing pseudospin operators. We want to notice that
the use of spin reflection positivity has been largely used in
literature to rigorously determine the symmetry character and the
correlation functions of the ground state for different model
Hamiltonians. In our work this idea is generalized and enriched in
a simple scheme which takes advantage of the positivity character
of the ground state in one region of the parameter space and of
the use of symmetry transformations to scan a larger part of the
phase diagram.
\\Moreover, the procedure presented above gives also
the opportunity to extract information on the phase diagram of the
model in exam, such as the occurrence of a novel quantum state
with coexistence of spin and orbital order, the manifestation of
ODLRO, and where the ground state can be a supersolid.
\\Finally, it is worth pointing out that a similar model has
been exactly solved in the one-dimensional case in the case of
strong coupling regime,{\cite{shen}} i.e. when the Coulomb
repulsion is such that the double occupancy of electrons on the
same site of the same orbital is excluded.

\end{multicols} %%%
\end{document}